# Graph Operations on Parity Games and Polynomial-Time Algorithms

Christoph Dittmann, Stephan Kreutzer, Alexandru I. Tomescu*

Chair for Logic and Semantics, Technical University Berlin
christoph.dittmann@tu-berlin.de, stephan.kreutzer@tu-berlin.de,
alexandru.tomescu@mailbox.tu-berlin.de

## 1 Introduction

Parity games are games that are played on directed graphs whose vertices are labeled by natural numbers, called priorities. The players push a token along the edges of the digraph. The winner is determined by the parity of the greatest priority occurring infinitely often in this infinite play.

A motivation for studying parity games comes from the area of formal verification of systems by model checking. Deciding the winner in a parity game is polynomial time equivalent to the model checking problem of the modal $\mu$-calculus (e.g., [4]). Another strong motivation lies in the fact that the exact complexity of solving parity games is a long-standing open problem, the currently best known algorithm being subexponential [7]. It is known that the problem is in the complexity class UP $\cap$ coUP [6].

In this paper we identify restricted classes of digraphs where the problem is solvable in polynomial time, following an approach from structural graph theory. We consider three standard graph operations: the join of two graphs, repeated pasting along vertices, and the addition of a vertex. Given a class $\mathcal{C}$ of digraphs on which we can solve parity games in polynomial time, we show that the same holds for the class obtained from $\mathcal{C}$ by applying once any of these three operations to its elements.

These results provide, in particular, polynomial time algorithms for parity games whose underlying graph is an orientation of a complete graph, a complete bipartite graph, a block graph, or a block-cactus graph. These are classes where the problem was not known to be efficiently solvable.

Previous results concerning restricted classes of parity games which are solvable in polynomial time include classes of bounded tree-width [9], bounded DAG-width [2], and bounded clique-width [10].

We also prove that recognising the winning regions of a parity game is not easier than computing them from scratch.

## 2 Preliminaries

A *directed graph* (digraph) is a pair $G = (V, E)$ where $V$ is the set of vertices and $E \subseteq V \times V$ is the set of arcs. For a vertex $v$, we write $N(v)$ for the set of neighbours of $v$, that is, $N(v) := \{u \in V \mid (u,v) \in E \text{ or } (v,u) \in E\}$. In this paper, a *graph* is always a directed graph and we only consider finite graphs. An *undirected graph* is a graph where $E$ is a symmetric relation. For undirected graphs, we write $\{x, y\} \in E$ instead of $(x, y) \in E$.

---

* The third author gratefully acknowledges the support of the European Science Foundation, activity "Games for Design and Verification".

Complying with the terminology in [1], we say that an *orientation* of an undirected graph $G = (V, E)$ is a digraph obtained from $G$ by replacing every edge $\{x, y\} \in E$ by one of the arcs $(x, y)$ or $(y, x)$, but not both. A *biorientation* of $G$ is a digraph obtained from $G$ by replacing every edge $\{x, y\} \in E$ with the arcs $(x, y)$ or $(y, x)$ or both.

## 3 Parity Games

### 3.1 Basic Definitions and Results

A *parity game* $P = (V, V_\circ, V_\square, E, \Omega)$ is a finite directed graph $(V, E)$ with a partitioning of the vertices $V = V_\circ \cup V_\square$ equipped with a priority map $\Omega : V \to \mathbb{N}$. We write $V(P)$ to denote the set of vertices of $P$, and similarly $V_\circ(P)$, $V_\square(P)$, $E(P)$ and $\Omega(P)$ to denote the respective parts of $P$.

A *play* on $P$ starts with a token placed on some vertex $v \in V$. If $v \in V_\circ$, Player $\circ$ moves the token to a successor of $v$, otherwise $V_\square$ moves it to a successor. If there is no successor, the respective player loses. If the play continues forever, Player $\circ$ wins the play if and only if the maximum priority that appears infinitely often is even.

A *positional strategy* for Player $\circ$ is a map $\rho : V_\circ \to V$ such that $\rho(v)$ is a successor of $v$ for all $v$ such that $v$ has a successor. A play $v = v_0, v_1, v_2, \ldots$ *conforms* to $\rho$ if $v_{i+1} = \rho(v_i)$ for all $i$ such that $v_i \in V_\circ$. A positional strategy $\rho$ is a *positional winning strategy* for Player $\circ$ from vertex $v$ if every play that starts at $v$ and conforms to $\rho$ is winning for Player $\circ$.

We call the set of vertices $W_\circ(P) \subseteq V$ from which Player $\circ$ has a positional winning strategy the *winning region* of Player $\circ$, similar for $W_\square$ and Player $\square$. We will write $W_\circ$, $W_\square$ if the game is clear from the context.

Parity games are *positionally determined* in the sense that $W_\circ \cup W_\square = V$ and $W_\circ \cap W_\square = \emptyset$ [4, Chapter 6]. For this reason, we only consider positional strategies from now on.

Given $A \subseteq V$, we denote by $P \cap A$ the parity game restricted to the vertices in $A$, that is, $(V \cap A, V_\circ \cap A, V_\square \cap A, E \cap (A \times A), \Omega\!\restriction_A)$. Similarly, we denote the game $P \cap (V \setminus A)$ by $P \setminus A$. Given a class of parity games $\mathcal{C}$, we say that $\mathcal{C}$ is *hereditary* if for all $P \in \mathcal{C}$ and all subsets $A$ of vertices of $P$, we have $P \cap A \in \mathcal{C}$. For a given game $P$, we call a game $P'$ a *proper subgame* of $P$ if $P' = P \cap V(P')$ and $P' \neq P$. For $d \in \mathbb{N}$, we denote by $\Omega^{-1}(d)$ the set of vertices having priority $d$.

The following two notions are well-known [4] and form the basis of the exponential-time algorithms of McNaughton [8] and Zielonka [11], and of the sub-exponential-time algorithm of Jurdziński, Paterson, Zwick [7].

For $i \in \{\circ, \square\}$, we denote by $\bar{i}$ the element of $\{\circ, \square\} \setminus \{i\}$. Given $i \in \{\circ, \square\}$, a set $A \subseteq V$ is said to be *i-closed* if for every $u \in A$, the following two properties hold:

- If $u \in V_i$, then there is some $(u, v) \in E$ such that $v \in A$.
- If $u \in V_{\bar{i}}$, then for every $(u, v) \in E$ we have $v \in A$.

We denote by $\mathrm{attr}_i(A)$ the set of vertices in $V$ from which Player $i$ has a strategy to enter $A$ at least once, and call it the *i-attractor set of* $A$. Clearly, the sets $W_i$, and $\mathrm{attr}_i(A)$, for every $A \subseteq V$, are $i$-closed, and $\mathrm{attr}_i(W_i) = W_i$. Attractor sets are particularly useful since they allow a decomposition of a parity game, as Lemmas 3.1 and 3.2 testify. Working with attractor sets is also computationally efficient, since they can be computed in time linear in the number of edges.

The following two lemmas are well-known results about $i$-closed sets and $i$-attractor sets.

**Lemma 3.1.** *Let $P = (V, V_\circ, V_\square, E, \Omega)$ be a parity game. For every $A \subseteq V$, and $i \in \{\circ, \square\}$, the set $V \setminus \mathrm{attr}_i(A)$ is $\bar{i}$-closed. Moreover, $W_{\bar{i}}(P \setminus \mathrm{attr}_i(A)) \subseteq W_{\bar{i}}(P)$.*

**Lemma 3.2.** *Let $P = (V, V_\circ, V_\square, E, \Omega)$ be a parity game and let $i \in \{\circ, \square\}$. If $U \subseteq W_i(P)$, then*

- $W_i(P) = \mathrm{attr}_i(U) \cup W_i(P \setminus \mathrm{attr}_i(U))$ and
- $W_{\bar{i}}(P) = W_{\bar{i}}(P \setminus \mathrm{attr}_i(U))$.

A *single-player game* is a parity game where all nodes belong to one of the players. We will use a function SOLVE-SINGLE-PLAYER-GAME($P$) which solves single-player games in time cubic in the number of vertices, by returning the two winning regions $(W_{\circ}(P), W_{\square}(P))$. It is easy to see that such an algorithm exists.

## 3.2 Half-Solving Parity Games

In the sequel we will repeatedly use the following lemma, whose proof is based on McNaughton's algorithm [8].

**Definition 3.3 (Half-Solving Parity Games).** *Let $\mathcal{C}$ be a class of parity games. An algorithm half-solving parity games in $\mathcal{C}$ is an algorithm which takes a game $P \in \mathcal{C}$ as input and returns one of the following three results.*

1. $W_{\circ}(P)$ and $W_{\square}(P)$.
2. *A proper subgame $P'$ of $P$ and sets $W_{\circ}^*, W_{\square}^* \subseteq V(P) \setminus V(P')$ such that $W_{\circ}(P) = W_{\circ}^* \cup W_{\circ}(P')$ and $W_{\square}(P) = W_{\square}^* \cup W_{\square}(P')$.*
3. *The game $P$ itself, but only if either $W_{\circ}(P) = \emptyset$ or $W_{\square}(P) = \emptyset$.*

**Lemma 3.4.** *Let $\mathcal{C}$ be a hereditary class of parity games. If there is an algorithm half-solving parity games in $\mathcal{C}$ running in time $O(n^c)$ for some $c > 0$, where $n$ is the number of vertices, then there is an algorithm that computes the winning regions on all parity games in $\mathcal{C}$ in time $O(n^{c+1})$.*

*Proof.* Let HALF-SOLVE be the given algorithm. Consider algorithm 1.

---

**Algorithm 1:** Turn a partial solution into a full solution.

SOLVE($P = (V, V_{\circ}, V_{\square}, E, \Omega)$)
$\quad R \leftarrow$ HALF-SOLVE($P$)
$\quad$ **if** $R = (W_{\circ}, W_{\square})$ **then**
$\quad\quad$ **return** $(W_{\circ}, W_{\square})$
$\quad$ **if** $R = (P', W_{\circ}^*, W_{\square}^*)$ **then**
$\quad\quad (W_{\circ}(P'), W_{\square}(P')) \leftarrow$ SOLVE($P'$)
$\quad\quad$ **return** $(W_{\circ}^* \cup W_{\circ}(P'), W_{\square}^* \cup W_{\square}(P'))$
$\quad$ **if** $R = P$ **then**
$\quad\quad d \leftarrow$ MAXIMUM-PRIORITY($\Omega$)
$\quad\quad i \leftarrow \circ$ **if** $d$ is even, $\square$ **otherwise**
$\quad\quad (C_{\circ}, C_{\square}) \leftarrow$ SOLVE($P \setminus \mathrm{attr}_i(\Omega^{-1}(d))$)
$\quad\quad$ **if** $C_{\bar{i}} \neq \emptyset$ **then**
$\quad\quad\quad (W_i, W_{\bar{i}}) \leftarrow (\emptyset, V)$
$\quad\quad$ **else**
$\quad\quad\quad (W_i, W_{\bar{i}}) \leftarrow (V, \emptyset)$
$\quad\quad$ **return** $(W_{\circ}, W_{\square})$

---

Let $T(n)$ be the running time of SOLVE and $kn^c$ be the running time of HALF-SOLVE with $k \in \mathbb{N}$. Then we have
$$T(n) \leq O(n) + kn^c + f(n) + T(n-1)$$

where $f(n)$ is the time to compute the attractor set $\text{attr}_i(\Omega^{-1}(d))$.

An attractor set can be computed in time $O(n^2)$, but this would only give the bound $T(n) \in O(n^{c+2})$. However, we observe that in total each arc is only considered once because $P \setminus \text{attr}_i(\Omega^{-1}(d))$ does not contain any arcs with their start or end point in $\text{attr}_i(\Omega^{-1}(d))$. So the algorithm only needs $O(n^2)$ time to compute all the attractor sets in total over all recursive steps. So we have $T(n) \in O(n^{c+1} + n^2)$ which gives the desired bound of $T(n) \in O(n^{c+1})$.

To prove the correctness, it is enough to consider the case where HALF-SOLVE returns $P$ unmodified because the other cases are trivially correct.

Let $D \subseteq V$ be the set of vertices of highest priority in $P$. Assume that their priority is good for Player $\circ$ (the case of Player $\square$ is similar). Then the algorithm solves the game $P' := P \setminus \text{attr}_\circ(D)$.

If $W_\square(P') \neq \emptyset$, then $W_\square(P) \neq \emptyset$ because every vertex winning for Player $\square$ in $P'$ is also winning for Player $\square$ in $P$, because we only removed a $\circ$-attractor (Lemma 3.1). Since we are in the case where one of the winning regions of $P$ is empty, this implies $W_\circ(P) = \emptyset$.

If $W_\square(P') = \emptyset$, then it is easy to see that Player $\circ$ wins everywhere in $P$. Indeed, if a play eventually stays in $P'$, then Player $\circ$ wins because $W_\square(P') = \emptyset$. If a play visits $\text{attr}_\circ(D)$ an infinite number of times, then Player $\circ$ can force to visit $D$ an infinite number of times, and hence wins the play. □

### 3.3 Recognising Winning Regions is Equivalent to Computing Them

In Lemma 3.4, we showed a method to decide the global winner under the restrictions that there is one definite global winner and under the assumption that we can solve subgames recursively. A natural question is if this can be generalised to recognise winning regions without this restriction.

More formally, given a parity game $P$ and a set $A \subseteq V(P)$, the problem of *recognising winning regions* is the problem to decide if $A = W_\square(P)$ holds.

A very specific special case of this problem is the problem of *recognising a global $\square$-winner*. This is the problem of deciding whether Player $\square$ wins on all nodes, that is, decide whether $W_\square(P) = V(P)$ holds.

Obviously, if we can compute the winning regions in polynomial time, we can also recognise them in polynomial time. Unfortunately, it also turns out that recognising simply a global $\square$-winner is just as hard as computing winning regions, as the following theorem shows.

**Theorem 3.5.** *If there is an algorithm that recognises a global $\square$-winner in polynomial time, then there is an algorithm that computes the winning regions of parity games in polynomial time.*

*Proof.* Assume that we can recognise a global $\square$-winner in time $O(n^c)$ for some constant $c$, where $n$ is the number of vertices, and let $P = (V, V_\circ, V_\square, E, \Omega)$ be a parity game with $n$ vertices.

Choose an arbitrary vertex $s \in V$. We are going to construct a game $P_s$ based on $P$ that tells us whether $s \in W_\circ(P)$ or $s \in W_\square(P)$. For all arcs $(u, v) \in E$ with $u \in V_\circ$, add a new disjoint $\square$-vertex $v_{u,v}$ to $P_s$ with the same priority as $u$ and replace the arc $(u, v)$ with the two arcs $(u, v_{u,v}), (v_{u,v}, v)$. Add a single $\circ$-vertex $w$ to $P_s$ and let $w$ have the unique priority that is worst for Player $\square$. Now add arcs $(u, w)$ for all $\square$-vertices $u$ to $P_s$. Finally, add the single arc $(w, s)$.

If $s \in W_\square(P)$, then $s \in W_\square(P_s)$ because Player $\square$ can choose to stay in the original game. But the $\square$-attractor of $s$ contains $w$ and hence all vertices in $P_s$, so then we have $W_\square(P_s) = V(P_s)$.

If $s \in W_\circ(P)$, then $s \in W_\circ(P_s)$ because Player $\square$ cannot move through $w$ infinitely often. If Player $\square$ only moves through $w$ finitely often, then he will eventually end up in a game in $P$ starting in $s$, which he loses by the assumption $s \in W_\circ(P)$. So we have $s \in W_\circ(P_s)$ and hence $W_\square(P_s) \neq V(P_s)$.

So by determining if $W_\square(P_s) = V(P_s)$ holds for each $s \in V$ in turn, we can construct the winning regions of the original game. The game $P_s$ has at most $n + n^2$ vertices, so in total this algorithm runs in time $O(n(n + n^2)^c) = O(n^{2c+1})$. □

# 4 The Join of Two Parity Games

In this section we will show that if $\mathcal{C}$ and $\mathcal{C}'$ are hereditary classes of parity games that can be solved in polynomial time, then the class of parity games obtained by *joining* games from $\mathcal{C}$ and $\mathcal{C}'$ can also be solved in parity games.

**Definition 4.1 (join of parity games).** *Given parity games $P' = (V', V'_\bigcirc, V'_\square, E', \Omega')$ and $P'' = (V'', V''_\bigcirc, V''_\square, E'', \Omega'')$ with $V' \cap V'' = \emptyset$, we say that the game $P = (V, V_\bigcirc, V_\square, E, \Omega)$ is a join of $P'$ and $P''$ if the following conditions hold:*

- $V = V' \cup V''$, $V_\bigcirc = V'_\bigcirc \cup V''_\bigcirc$, $V_\square = V'_\square \cup V''_\square$,
- $E = E' \cup E'' \cup E^*$, where $E^* \subseteq (V' \times V'') \cup (V'' \times V')$ contains at least one arc $(x,y)$ or $(y,x)$ for all $x \in V'_\square$, $y \in V''_\bigcirc$ and for all $x \in V'_\bigcirc$, $y \in V''_\square$,
- *the vertices of $P$ have the same priorities as they have in $P'$ and $P''$.*

*Given two classes of parity games $\mathcal{C}$ and $\mathcal{C}'$, we define*

$$\mathsf{HalfJoin}(\mathcal{C}) := \{P \mid P \text{ is a join of a single-player game } P' \text{ and a game } P'' \in \mathcal{C}\},$$
$$\mathsf{Join}(\mathcal{C}, \mathcal{C}') := \{P \mid P \text{ is a join of } P' \in \mathcal{C} \text{ and } P'' \in \mathcal{C}'\}.$$

*Remark 4.2.* Observe that if $\mathcal{C}$ and $\mathcal{C}'$ are hereditary, then so are $\mathsf{HalfJoin}(\mathcal{C})$ and $\mathsf{Join}(\mathcal{C}, \mathcal{C}')$.

We will first show how to solve parity games obtained by joining a polynomial time solvable parity game with a single-player game and then extend this construction to the general case of joining arbitrary parity games.

As an immediate corollary we get that parity games whose underlying undirected graph is a complete graph, so-called *tournaments*, can be solved in polynomial time. As corollary from the more general case of arbitrary joins we get that parity games whose underlying undirected graph is a complete bipartite graph can be solved in polynomial time. Note that the result for tournaments is not a special case of Obdržálek's polynomial time algorithm [10] for parity games of bounded directed clique-width because biorientations of complete graphs and of complete bipartite graphs do not have bounded directed clique-width, although their underlying undirected graphs have bounded undirected clique-width.

## 4.1 Adjoining Vertices Belonging to One Player

We now define the construction that joins a parity game $P$ and a parity game $P'$ whose vertices belong to only one player. This construction will also be used later in the proof of the main theorem on joins, given in Section 4.2.

**Theorem 4.3.** *If $\mathcal{C}$ is a hereditary class of parity games that can be solved in polynomial time, then all games $P \in \mathsf{HalfJoin}(\mathcal{C})$ can be solved in polynomial time, provided that a decomposition of $P$ as a join of a game in $\mathcal{C}$ with a single-player parity game is given.*

*Proof.* Let $P = (V, V_\bigcirc, V_\square, E, \Omega) \in \mathsf{HalfJoin}(\mathcal{C})$ be a join of the single-player game $P' = (V', V'_\bigcirc, V'_\square, E', \Omega')$ and the game $P'' = (V'', V''_\bigcirc, V''_\square, E'', \Omega'') \in \mathcal{C}$. We may assume without loss of generality that $V'_\bigcirc$ is empty, so that $V' = V'_\square$ and $V_\bigcirc = V''_\bigcirc$.

Let $O(n^c)$ be the time complexity for solving games in $\mathcal{C}$ where $n$ is the number of vertices. Recall that single-player parity games can be solved in cubic time by the algorithm SOLVE-SINGLE-PLAYER-GAME($P$).

**Algorithm 2:** A polynomial-time algorithm for solving parity games on half joins where $P$ is a join the single-player game $P'$ and the game $P'' \in \mathcal{C}$.

HALF-SOLVE-HALF-JOIN($P = (V, V_\circ, V_\square, E, \Omega)$, $P'$, $P''$)
    **if** $V = \emptyset$ **then**
        **return** $(\emptyset, \emptyset)$
    $i \leftarrow \circ$ **if** $V_\circ(P') = \emptyset$, $\square$ **otherwise**
    $(A_i, A_{\bar{i}}) \leftarrow$ SOLVE-SINGLE-PLAYER-GAME($P \setminus \text{attr}_i(V_i(P''))$)
    **if** $A_{\bar{i}} \neq \emptyset$ **then**
        $W_i^* \leftarrow \emptyset$
        $W_{\bar{i}}^* \leftarrow \text{attr}_{\bar{i}}(A_{\bar{i}})$
        **return** $(P \setminus W_{\bar{i}}^*, W_\circ^*, W_\square^*)$
    $(B_i, B_{\bar{i}}) \leftarrow$ SOLVE-$\mathcal{C}$-GAME($P \setminus \text{attr}_{\bar{i}}(V_{\bar{i}}(P'))$)
    **if** $B_i \neq \emptyset$ **then**
        $W_i^* \leftarrow \text{attr}_i(B_i)$
        $W_{\bar{i}}^* \leftarrow \emptyset$
        **return** $(P \setminus W_i^*, W_\circ^*, W_\square^*)$
    **return** $P$

By Lemma 3.4 it suffices to establish an algorithm half-solving the parity games in $\mathsf{HalfJoin}(\mathcal{C})$. See algorithm 2 for a detailed description of our algorithm.

The algorithm first solves the single-player game $P_1 := P \setminus \text{attr}_\circ(V''_\circ)$, in time $O(n^3)$. See Figure 1(a) for an illustration of the situation, where nodes lie above the dotted line if and only if they are winning for Player $\square$. By Lemma 3.1 we know that $W_\square(P_1) \subseteq W_\square(P)$ and by Lemma 3.2 we get that

$$W_\square(P) = \text{attr}_\square(W_\square(P_1)) \cup W_\square(P \setminus \text{attr}_\square(W_\square(P_1))).$$

Therefore, if $W_\square(P_1)$ is not empty, then we can return $(P \setminus \text{attr}_\square(W_\square(P_1)), \emptyset, \text{attr}_\square(W_\square(P_1)))$, which is the second outcome of the algorithm according to Definition 3.3.

If $W_\square(P_1)$ is empty, then we solve the game $P_2 := P \setminus \text{attr}_\square(V'_\square)$, in time $O(n^c)$, and proceed as before (see Figure 1(b)). By Lemma 3.1 we know that $W_\circ(P_2) \subseteq W_\circ(P)$. Thus, if $W_\circ(P_2)$ is not empty, then we can return $(P \setminus \text{attr}_\circ(W_\circ(P_2)), \text{attr}_\circ(W_\circ(P_2)), \emptyset)$, which again is the second possible outcome according to Definition 3.3.

Finally, suppose both $W_\square(P_1)$ and $W_\circ(P_2)$ are empty. We claim that either $W_\square(P) = \emptyset$ or $W_\circ(P) = \emptyset$. In both cases we return $P$ unmodified as this is the third possible outcome according to Definition 3.3. To establish the claim we distinguish two cases, depending on whether $V'_\square \cap W_\circ(P) = \emptyset$ or not.

**Case 1.** $V'_\square \cap W_\circ(P) \neq \emptyset$ (see Figure 2(a)).
  Observe that

$$V''_\circ \subseteq W_\circ(P).$$

Assume to the contrary that there is $v \in V''_\circ \cap W_\square(P)$ and let $w \in V'_\square \cap W_\circ(P)$. But then there can be no arc $(v, w)$ in $P$ because if there were such an arc, then Player $\circ$ would have a winning strategy from $v$ by choosing $w$ as $v$'s successor. Similarly, there cannot be an arc $(w, v)$. This contradicts the definition of $P$. By Lemma 3.2, we have that

$$W_\square(P) = W_\square(P \setminus \text{attr}_\circ(V''_\circ)) = W_\square(P_1).$$

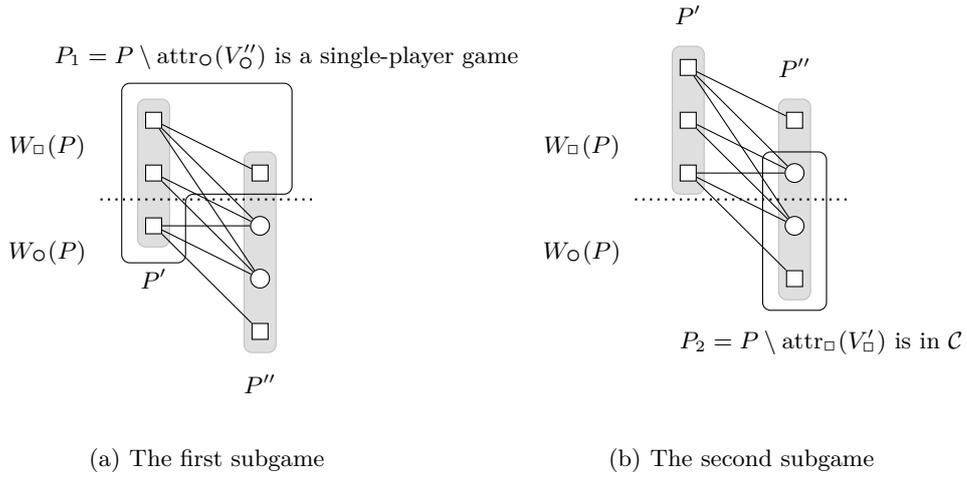

(a) The first subgame  (b) The second subgame

**Fig. 1.** An illustration of the subgames solved by algorithm 2, where $P$ is a join of the single-player parity game $P'$ and the game $P'' \in \mathcal{C}$.

As $W_\square(P_1) = \emptyset$, this implies that $W_\bigcirc(P) = V$.

**Case 2.** $V'_\square \cap W_\bigcirc(P) = \emptyset$, equivalently $V'_\square \subseteq W_\square(P)$ (see Figure 2(b)). Again by Lemma 3.2, we have that

$$W_\bigcirc(P) = W_\bigcirc(P \setminus \mathrm{attr}_\square(V'_\square)) = W_\bigcirc(P_2).$$

As $W_\bigcirc(P_2) = \emptyset$, this implies that $W_\square(P) = V$.

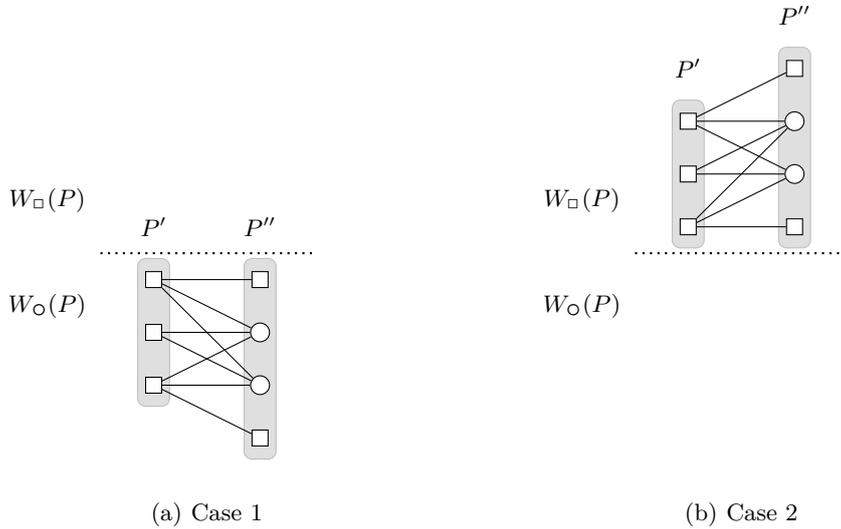

(a) Case 1  (b) Case 2

**Fig. 2.** An illustration of the case distinction of the proof of Theorem 4.3, where $P$ is a join of the single-player parity game $P'$ and the game $P'' \in \mathcal{C}$.

We see that the running time of the algorithm is $O(n^3 + n^c)$ because solving the single-player game and the game in $\mathcal{C}$ are the most expensive operations. With Lemma 3.4, we get an algorithm that solves all games in $\mathsf{HalfJoin}(\mathcal{C})$ in time $O(n^{1+\max\{3,c\}})$. □

**Definition 4.4.** *We say that a digraph $D = (V, E)$, with a partition of its vertices $V = V_\circ \cup V_\square$, is a weak tournament if between every two vertices $v \in V_\circ$, $w \in V_\square$ we have that $(v, w) \in E$ or $(w, v) \in E$ (or both).*

*We denote by* $\mathsf{wTournaments}$ *the class of parity games on a weak tournament, that is,*

$$\mathsf{wTournaments} := \{P = (V, V_\circ, V_\square, E, \Omega) \mid (V, E) \text{ with the partition } V = V_\circ \cup V_\square$$
$$\text{is a weak tournament}\}.$$

An easy corollary of Theorem 4.3 is the following.

**Corollary 4.5.** *There is an algorithm that solves all parity games $P = (V, V_\circ, V_\square, E, \Omega) \in \mathsf{wTournaments}$ and runs in time $O(|V|^4)$.*

*Proof.* Let $\mathcal{C}$ be the class of single-player games. Then $\mathsf{HalfJoin}(\mathcal{C}) = \mathsf{wTournaments}$. □

For $i \in \{\circ, \square\}$, we define $\mathsf{HalfJoin}_i(\mathcal{C})$ to be the class of joins of single-player games that belong to player $i$ with games from $\mathcal{C}$. Note that $\mathsf{HalfJoin}(\mathcal{C}) = \mathsf{HalfJoin}_\circ(\mathcal{C}) \cup \mathsf{HalfJoin}_\square(\mathcal{C})$. We call a class $\mathcal{C}$ of parity games *polynomial-time decidable* if there is a polynomial-time algorithm that decides $\mathcal{C}$, that is, tests membership $P \in \mathcal{C}$.

We see that in some cases we do not need to provide a decomposition as required by Theorem 4.3.

**Corollary 4.6.** *Let $\mathcal{C}$ be a hereditary class of parity games and $i \in \{\circ, \square\}$.*
1. *If $\mathcal{C}$ is polynomial-time solvable, then $\mathsf{HalfJoin}_i(\mathcal{C})$ is polynomial-time solvable.*
2. *If $\mathcal{C}$ is polynomial-time solvable and polynomial-time decidable, then $\mathsf{HalfJoin}(\mathcal{C})$ is polynomial-time solvable.*

*Remark 4.7.* Given $P = (V, V_\circ, V_\square, E, \Omega) \in \mathsf{HalfJoin}_i(\mathcal{C})$, we can compute a decomposition of it in linear time. Indeed, we can take

- $P'_i := P \cap \{v \in V_i \mid N(v) = V_{\bar{i}}\}$ as the single-player game,
- $P''_i := P \setminus \{v \in V_i \mid N(v) = V_{\bar{i}}\}$ as the game from the hereditary class $\mathcal{C}$.

This proves the first statement.

If the player to which the vertices of the single-player game belong to is not known, then we have to check for both $i \in \{\circ, \square\}$ which game $P''_i$ considered above belongs to $\mathcal{C}$ in order to find a decomposition. This proves the second statement.

Even though not needed in the next section, we introduce here another kind of adjoining operation between a single-player parity game $P'$ and an arbitrary parity game $P''$. In this case, assuming the vertices of $P'$ belong to Player $\bar{i}$, we fix a subset $M$ of vertices of Player $i$ of $P''$ and connected every vertex of $P'$ with every vertex in $M$.

**Definition 4.8.** *Given $i \in \{\square, \circ\}$, a single-player game $P' = (V', V'_\circ, V'_\square, E', \Omega')$ with $V'_{\bar{i}} = \emptyset$ and a parity game $P'' = (V'', V''_\circ, V''_\square, E'', \Omega'')$ with $V' \cap V'' = \emptyset$, we say that a game $P = (V, V_\circ, V_\square, E, \Omega)$ is a generalised single-player join (G-join) of $P'$ and $P''$ if $P$ is the join of $P'$ and $P''$ except that the set of edges is defined differently: There exists a set $M \subseteq V''_i$ such that $E = E' \cup E'' \cup E^*$ and $E^* \subseteq (M \times V'_{\bar{i}}) \cup (V'_{\bar{i}} \times M)$ and there is at least one arc $(x, y)$ or $(y, x)$ for all $x \in M$, $y \in V'_{\bar{i}}$.*

*If $\mathcal{C}$ is a class of parity games, then we denote by $\mathsf{HalfJoin}_G(\mathcal{C})$ the class*

$$\mathsf{HalfJoin}_G(\mathcal{C}) := \{P \mid P \text{ is a G-join of a single-player game } P' \text{ and a game } P'' \in \mathcal{C}\}.$$

**Theorem 4.9.** *If $\mathcal{C}$ is a hereditary class of parity games that can be solved in polynomial time, then all games $P \in \mathsf{HalfJoin}_G(\mathcal{C})$ can be solved in polynomial time, provided that a decomposition of $P$ as a $G$-join of a single-player game and a game in $\mathcal{C}$ is given.*

*Proof.* The proof is very similar to the proof of Theorem 4.3, the only difference being in Case 1. We briefly sketch this difference here, under the same notations and assumptions as before.

If the set $V'_\square \cap W_\bigcirc(P)$ is not empty, then let $v$ be an arbitrary vertex in this set, and let $M := N(v) \cap V''_\bigcirc$. By the same argument given before, $M \subseteq W_\bigcirc(P)$. Therefore, there is no arc between a vertex in $W_\square(P) \cap V'$ and a vertex of $W_\square(P) \cap V''$ because $E \subseteq E' \cup E'' \cup (M \times V''_\bigcirc) \cup (V''_\bigcirc \times M)$. Thus, $W_\square(P) = W_\square(P') \cup W_\square(P'')$, both $W_\square(P')$ and $W_\square(P'')$ being computable in polynomial time. If $W_\square(P) \neq \emptyset$, then we can remove $W_\square(P)$ from the game by Lemma 3.2 (Case 1a). Otherwise, $W_\square(P) = \emptyset$, implying that $W_\bigcirc(P) = V$ (Case 1b).

The complementary case $V'_\square \cap W_\bigcirc(P) = \emptyset$ is identical to Case 2 in the proof of Theorem 4.3, and the result immediately follows. □

### 4.2 Joining Two Parity Games

Now we can state our main theorem on the join of two parity games.

**Theorem 4.10.** *If $\mathcal{C}$ and $\mathcal{C}'$ are hereditary classes of parity games that we can solve in polynomial time, then there is an algorithm for solving parity games in polynomial time on all games $P \in \mathsf{Join}(\mathcal{C}, \mathcal{C}')$, assuming a decomposition of $P$ as a join of a game in $\mathcal{C}$ and a game in $\mathcal{C}'$ is given.*

*Proof.* Let $P = (V, V_\bigcirc, V_\square, E, \Omega) \in \mathsf{Join}(\mathcal{C}, \mathcal{C}')$ be a join of a game $P' = (V', V'_\bigcirc, V'_\square, E', \Omega') \in \mathcal{C}$ with a game $P'' = (V'', V''_\bigcirc, V''_\square, E'', \Omega'') \in \mathcal{C}'$. Let $O(n^{c_1})$ and $O(n^{c_2})$ be the time complexities for solving games in $\mathcal{C}$ and in $\mathcal{C}'$, respectively, where $n$ is the number of nodes.

We follow the lines of the proof of Theorem 4.3. The case distinctions and their conclusions are exactly the same. The only difference is that in the proof of Theorem 4.3, the two subgames $P \setminus \mathsf{attr}_\bigcirc(V''_\bigcirc)$ and $P \setminus \mathsf{attr}_\square(V'_\square)$ were solvable in polynomial time because they were single-player games or in $\mathcal{C}$, respectively. Here these games can be solved in time $O(n^{1+\max\{3,c_1,c_2\}})$ by Theorem 4.3 because both games are in $\mathsf{HalfJoin}(\mathcal{C})$ and $\mathsf{HalfJoin}(\mathcal{C}')$, respectively. See Figure 3 and note the similarity to Figures 1(a) and 1(b).

The remaining part of the algorithm is exactly the same as in the proof of Theorem 4.3: If both of these subgames do not provide a usable winning region, then we have $W_\square(P) = \emptyset$ or $W_\bigcirc(P) = \emptyset$ and we return the original game unmodified.

In total this gives a running time of $O(n^{1+\max\{3,c_1,c_2\}})$ for the preliminary algorithm. Lemma 3.4 turns this into an algorithm that solves all parity games in $\mathsf{Join}(\mathcal{C}, \mathcal{C}')$ in time $O(n^{2+\max\{3,c_1,c_2\}})$. □

We remark that if $\mathcal{C}$ is the class of parity games without arcs, then the class $\mathsf{Join}(\mathcal{C}, \mathcal{C})$ is the class of parity games whose underlying graph is a biorientation of a complete bipartite graph.

## 5 Pasting of Parity Games

**Definition 5.1.** *Let $P' = (V', V'_\bigcirc, V'_\square, E', \Omega')$ and $P'' = (V'', V''_\bigcirc, V''_\square, E'', \Omega'')$ be two parity games with $V' \cap V'' = \emptyset$, and let $v' \in V'$ and $v'' \in V''$. Assume that $v', v''$ have the same priority and belong to the same player in $P'$, and in $P''$, respectively. We say that the game $P = (V, V_\bigcirc, V_\square, E, \Omega)$ is obtained by pasting $P', P''$ at $v', v''$, if $P$ is the disjoint copy of $P'$ and $P''$ with $v', v''$ identified (see Figure 4(a)). Formally, assuming that $v', v''$ belong to Player $i$,*

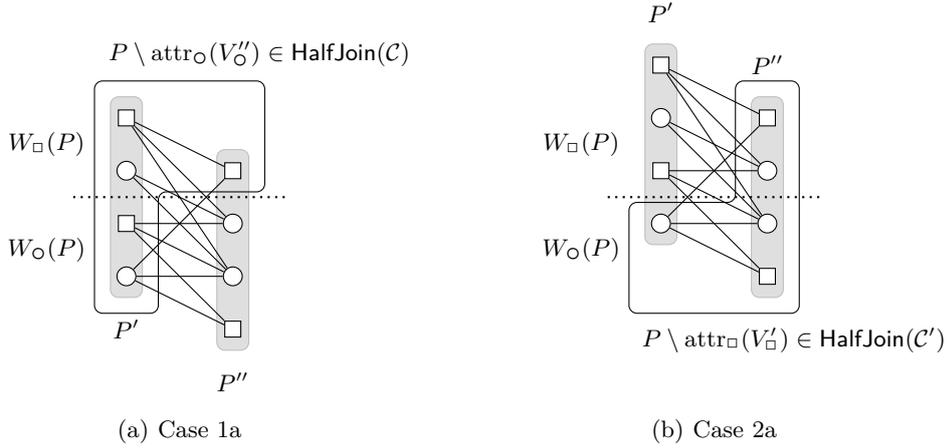

(a) Case 1a  (b) Case 2a

**Fig. 3.** An illustration the proof of Theorem 4.10, where $P$ is a join of $P' \in \mathcal{C}$ and $P'' \in \mathcal{C}'$.

- $V = (V' \cup V'' \cup \{v^*\}) \setminus \{v', v''\}$, $V_i = (V'_i \cup V''_i \cup \{v^*\}) \setminus \{v', v''\}$, $V_{\bar{i}} = V'_{\bar{i}} \cup V''_{\bar{i}}$,
- $E = \{(u, w) \in E' \cup E'' \mid \{v', v''\} \cap \{u, w\} = \emptyset\}) \cup$
  $\{(u, v^*) \mid (u, v') \in E' \vee (u, v'') \in E''\} \cup$
  $\{(v^*, u) \mid (v', u) \in E' \vee (v'', u) \in E''\}$,
- the vertices of $P$ have the same priorities as they have in $P'$ and $P''$, where $\Omega(v^*) = \Omega(v') = \Omega(v'')$.

Given a class of parity games $\mathcal{C}$, we denote by $\mathsf{RepeatedPasting}(\mathcal{C})$ the class of games obtained by repeated pasting of a finite number of games from $\mathcal{C}$.

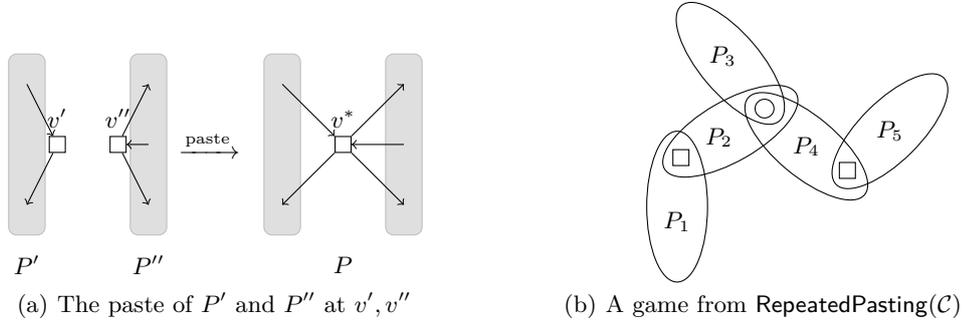

(a) The paste of $P'$ and $P''$ at $v', v''$  (b) A game from $\mathsf{RepeatedPasting}(\mathcal{C})$

**Fig. 4.** The paste operation.

We observe that if $\mathcal{C}$ is hereditary and is closed under disjoint unions, then $\mathsf{RepeatedPasting}(\mathcal{C})$ is hereditary. Moreover, every $P \in \mathsf{RepeatedPasting}(\mathcal{C})$ has a decomposition into components $P_1, P_2, \ldots, P_k$ with $P_i \in \mathcal{C}$, such that every distinct $P_i$ and $P_j$ are either disjoint or share exactly one vertex. This means that games $P_1, \ldots, P_k$ form a tree-like structure, in the sense that the graph $T_P$ obtained by adding a vertex $i$ for every $P_i$ and an edge $(i, j)$ if $P_i$ and $P_j$ share a vertex is a tree (see Figure 4(b)).

**Theorem 5.2.** *If $\mathcal{C}$ is a hereditary class of parity games that can be solved in time $O(n^c)$, then games in* RepeatedPasting$(\mathcal{C})$ *can be solved in time $O(n^{1+\max\{2,c\}})$.*

*Proof.* Let $P \in$ RepeatedPasting$(\mathcal{C})$ and let $T(n)$ be the running time of the algorithm we are constructing for RepeatedPasting$(\mathcal{C})$, where $n$ is the number of vertices of $P$. In order to find a decomposition of $P$ into games from $\mathcal{C}$, we compute the biconnected components (that is, maximal 2-connected subgraphs) of the underlying undirected graph of $P$, say $P_1, \ldots, P_k$, in time linear in the number of edges of $P$, for example by the algorithm presented in [5]. Since $P \in$ RepeatedPasting$(\mathcal{C})$ and $\mathcal{C}$ is hereditary, each such biconnected component belongs to $\mathcal{C}$.

Let $L \in \mathcal{C}$ be a leaf-component of the tree $T_P$ associated with $P$. This means that $L$ shares at most one vertex with all other components of $P$. If there is no such vertex, we are done because the graph is disconnected and we can easily solve different components separately. Otherwise, let $v$ be this vertex. Without loss of generality we assume that $v$ is a vertex that belongs to Player $\circ$. Since all paths between $L$ and $P' := P \setminus (L \setminus \{v\})$ use vertex $v$, and thus there are no cycles without repeated vertices spanning $L$ and $P'$, we have that $v \in W_\circ(P)$ if and only if $v \in W_\circ(L)$ or $v \in W_\circ(P \setminus (L \setminus \{v\}))$.

First we solve the parity game $L$, which can be done in time $O(n^c)$ since $L \in \mathcal{C}$. If Player $\circ$ wins on $v$ in $L$, then we solve $P \setminus (L \cup \text{attr}_\circ(v))$ recursively, in time $T(n-1)$. Since all paths between $L$ and $P \setminus (L \cup \text{attr}_\circ(v))$ use vertex $v$, from which Player $\circ$ has a winning strategy inside $L$, we have, thanks to Lemma 3.2:

- $W_\circ(P) = W_\circ(L) \cup \text{attr}_\circ(v) \cup W_\circ(P \setminus (L \cup \text{attr}_\circ(v)))$,
- $W_\square(P) = W_\square(L) \cup W_\circ(P \setminus (L \cup \text{attr}_\circ(v)))$.

However, if Player $\square$ wins on $v$ in $L$, then we solve $P'$ recursively, in time $T(n-1)$. Then we have two cases. If Player $\square$ wins on $v$ in $P'$, then we merge the winning regions of $L$ and $P'$, that is:

- $W_\circ(P) = W_\circ(L) \cup W_\circ(P')$,
- $W_\square(P) = W_\square(L) \cup W_\square(P')$.

Otherwise, if Player $\circ$ wins on $v$ in $P'$, then we have to re-compute the winning regions of $L$, from which we remove $\text{attr}_\circ(v)$. Since $\mathcal{C}$ is hereditary, then $L \setminus \text{attr}_\circ(v) \in \mathcal{C}$, and can be solved in time $O(n^c)$. Thus, again by Lemma 3.2, and by the fact that all paths between $L \setminus \text{attr}_\circ(v)$ and $P'$ use vertex $v$, from which Player $\circ$ has a winning strategy inside $P'$, we put:

- $W_\circ(P) = W_\circ(L \setminus \text{attr}_\circ(v)) \cup \text{attr}_\circ(v) \cup W_\circ(P')$,
- $W_\square(P) = W_\square(L \setminus \text{attr}_\circ(v)) \cup W_\square(P')$.

Finally, the running time of the algorithm satisfies

$$T(n) \leq n^2 + tn^c + T(n-1),$$

for some fixed $t > 1$, where $n^2$ accounts for the computation of the attractor sets. Thus, $T(n) \in O(n^{1+\max\{2,c\}})$. □

As a corollary of Corollary 4.5 and Theorem 5.2, we can solve parity games in polynomial time on any biorientation of a *block-cactus graph* [3], that is, a graph whose biconnected components are cliques or cycles.

On the other hand, it is unlikely that we can easily extend the above method by pasting along 2 vertices since this immediately leads to a polynomial-time algorithm for the class of all parity games by pasting along complete graphs with 2 vertices.

## 6 Adding a Single Vertex

**Definition 6.1.** *If $\mathcal{C}$ is a class of parity games, we let $\mathsf{AddVertex}(\mathcal{C})$ denote the class of parity games obtained by adding a single vertex to every game in $\mathcal{C}$ in any possible way. Formally,*

$$\mathsf{AddVertex}(\mathcal{C}) := \{P \mid P \text{ is a parity game and there exists a vertex } v \text{ such that } P \setminus \{v\} \in \mathcal{C}\}.$$

**Theorem 6.2.** *If $\mathcal{C}$ is a hereditary class of parity games which we can solve in time $O(n^c)$ with $c \geq 2$ and we can test membership to $\mathcal{C}$ in time $O(n^d)$, then we can solve parity games in $\mathsf{AddVertex}(\mathcal{C})$ in time $O(n^{c+1})$ and test membership to $\mathsf{AddVertex}(\mathcal{C})$ in time $O(n^{d+1})$, where $n$ is the number of vertices of a parity game.*

*Proof.* Let $P = (V, V_\circ, V_\square, E, \Omega)$. In order to find a vertex $v$ such that $P \setminus \{v\} \in \mathcal{C}$, and at the same time test whether $P \in \mathsf{AddVertex}(\mathcal{C})$, we can iterate over all $v \in V$ and test the membership of $P \setminus \{v\}$ to $\mathcal{C}$, with an overall complexity of $O(n^{d+1})$.

First, we solve the two subgames $P_1 := P \setminus \mathsf{attr}_\circ(v) \in \mathcal{C}$ and $P_2 := P \setminus \mathsf{attr}_\square(v) \in \mathcal{C}$ in time $O(n^c)$. If $W_\square(P_1)$ or $W_\circ(P_2)$ is not empty, we return the corresponding subgame $P \setminus P_1$ or $P \setminus P_2$.

If $W_\square(P_1) = W_\circ(P_2) = \emptyset$, we conclude that $W_\square(P) = \emptyset$ or $W_\circ(P) = \emptyset$, analogous to the proof of theorems 4.3 and 4.10. We can then return the original game unchanged.

The running time of the whole algorithm is $O(n^c)$ because computing the attractor sets is in $O(n^2) \subseteq O(n^c)$. Lemma 3.4 then yields an algorithm that solves all games in $\mathsf{AddVertex}(\mathcal{C})$ in time $O(n^{c+1})$. □

This theorem implies, for example, that if parity games can be solved in polynomial time on orientations planar graphs, then they can also be solved in polynomial time on orientations apex graphs, which are planar graphs with one additional vertex.

## 7 Conclusions

We presented some graph operations that preserve solvability of parity games in polynomial time. In Section 4.2, we saw that the join of two classes of parity games is as easy to solve as the individual classes up to a polynomial factor. In Section 5 we considered the case of pasting many games together along vertices to form a larger game and in Section 6 we analysed the problem of adding a single vertex to a parity game. In both cases we showed that the resulting classes can be solved in time only a small polynomial factor slower than the original classes.

It is an open problem whether our approach can be adapted to further graph operations. One graph operations that comes to mind is the operation of *substitution*. In particular, let $T$ be an undirected tree where each node is coloured with either ○ or □ such that no two adjacent nodes have the same color. Then replace every $i$-coloured node with a single-player game consisting of nodes of player $i$ and connect games that correspond to adjacent nodes in $T$ analogous to the Join-operation defined in Section 4. The resulting game can be solved with dynamic programming and the help of Theorem 4.10 in time $O(n^c)$, but $c$ depends on the depth of the tree $T$. These games seem to be simpler than general parity games, so it could be reasonable to expect a polynomial time algorithm.